# Cotunnite-structured titanium dioxide: the hardest known oxide

**Leonid S. Dubrovinsky**[*], **Natalia A. Dubrovinskaia**[*], **Varghese Swamy**[†], **Joseph Muscat**[†], **Nicholes M. Harrison**[‡], **Rajeev Ahuja**[§] **& Borja Holm**[§]

[*] *Institute of Earth Sciences, Uppsala University, S-752 36 Uppsala, Sweden*

[†] *CSIRO Minerals, Box 312, Clayton South, Victoria 3169, Australia*

[‡] *CCLRC Daresbury Laboratory, Daresbury, Warrington, WA4 4AD, UK*
  *and Department of Chemistry, Imperial College London, London, SW7 2AY, UK*

[§] *Department of Physics, Uppsala University, S-751 21 Uppsala, Sweden*

**Despite great technological importance and many investigations[1-6], a material with measured hardness comparable to that of diamond or cubic boron nitride has yet to be identified. Combined theoretical and experimental investigations led to the discovery of a new polymorph of titanium dioxide with titanium nine-coordinated to oxygen in the cotunnite ($PbCl_2$) structure. Hardness measurements on the cotunnite-structured $TiO_2$ synthesized at pressures above 60 GPa and temperatures above 1000 K reveal that this material is the hardest oxide yet discovered. Furthermore, it is one of the least compressible (with a measured bulk modulus of 431 GPa) and hardest (with a microhardness of 38 GPa) polycrystalline materials studied thus far.**

It has been recognized that the hardness (H) of ionic and covalent materials is related to elastic properties and increases with bulk modulus ($K_T$) and shear modulus (G)[1,5,7-9] (Table 1). Diamond has the highest observed bulk (444 GPa) and shear (535 GPa) moduli. Diamond is also the hardest material known, with hardness values in the range of 90-100 GPa for single crystal and 50 GPa for sintered diamond[7,8,10]. It is followed by cubic boron nitride (cBN), with corresponding values of $K_{300}$ ($K_T$ at T=300 K) of 369 GPa and G of 409 GPa, and single-crystal H of 48 GPa and polycrystalline H of 32[7,8,10]. Stishovite, the high-pressure form of $SiO_2$, is the hardest oxide reported to date with an H of 33 GPa and a $K_{300}$ of 291 GPa[7,11].

A number of experimental and theoretical studies[12-15] indicate that titanium dioxide could have a series of high-pressure phases with their hardness possibly approaching that of diamond. For example, the baddeleyite-type TiO$_2$ has a K$_{300}$ of 290 GPa[14], a value that is close to the bulk modulus of stishovite. The monoclinic baddeleyite-type structure (MI, space group *P2$_1$/c*) is common among seven-fold coordinated dioxides and is known to transform, upon further compression, through an intermediate orthorhombic (OI, space group *Pbca*) structure to another orthorhombic (OII, space group *Pnma*) cotunnite-type phase[6]. Cotunnite-structured ZrO$_2$ and HfO$_2$ possess extremely high bulk moduli of 444 and 340 GPa, respectively[6]. If TiO$_2$ could also exist in the cotunnite structure, one can expect that such a material will be very incompressible and hard.

Previous *ab initio* simulations have successfully modeled the structural, elastic, and energetic properties of a number of well characterized TiO$_2$ phases including rutile, anatase, and the columbite-type (space group *Pbcn*) TiO$_2$-II[15,16]. In this study, we performed lattice dynamic (LD) and *ab initio* simulations (linear combination of atomic orbital periodic Hartree-Fock (LCAO-HF) and Full-Potential Linear-Muffin-Tin Orbital (FPLMTO) -- see *Methods*) of reasonable structures at pressures to 100 GPa to identify possible structures that TiO$_2$ could adopt under increased pressures. For the *ab initio* simulations, we have chosen very strict computational parameters that have been shown to yield reliable results for the bulk phases and surfaces of TiO$_2$[16]. The various structures simulated are rutile, anatase, TiO$_2$-II, MI, pyrite (Pa$\bar{3}$), fluorite (Fm3m), OI, and OII[13-16]. The simulations predict that the cotunnite-structured phase of TiO$_2$ is more stable than other structures proposed to date including the fluorite and pyrite modifications above 50 GPa (Fig. 1). Significantly, the *ab initio* calculations predict a remarkably high value for the bulk modulus for this phase: 380(20) GPa by LCAO-HF and 386(10) GPa by FPLMTO calculations.

We conducted a series of experiments in laser- or electric-heated diamond anvil cells (DACs) (see *Methods*) in order to determine whether or not it is possible to synthesize cotunnite-structured TiO$_2$. Anatase or rutile (99.99% TiO$_2$) was used as the starting material. At applied pressures of about 12 GPa or above, both rutile and anatase transformed to the baddeleyite (*P2$_1$/c*) phase, in good agreement with previous observations[12-14] (Figs. 2 and 3). On further compression, reflections due to the MI phase could be followed to over 60 GPa. Unit cell parameters of the MI phase were determined at pressures of 15 to 42 GPa and the molar volume (V) versus pressure (P) data were fitted to a third-order Birch-Murnaghan equation of state[17] (EOS):

$$P = 1.5\, K_{300}\, [(V_0/V)^{7/3} - (V_0/V)^{5/3}]\, [1 - 0.75\, (4-K')\{(V_0/V)^{2/3} - 1\}] \qquad (1)$$

where $K_{300}$, $K'$, and $V_0$ are the bulk modulus, its pressure derivative, and the molar volume at zero pressure and 300 K temperature, respectively. This yielded values of $K_{300}$ and $K'$ of 304(6) and 3.9(2) respectively, in good agreement with the values reported by Olsen et al.[14] ($K_{300}$=290(20) GPa at a fixed $K'$=4). At pressures above about 45 GPa, however, the quality of the diffraction pattern decreased drastically and at about 60 GPa, the material became translucent. We observed that at pressures above 50 GPa, $TiO_2$ absorbs Nd:YAG laser radiation and the laser-heated areas of the sample became black. After heating at 1600-1800 K by laser for 40 minutes at pressures between 60 and 65 GPa, the material transformed to a new phase as evidenced by the X-ray diffraction spectra (Figs. 2 and 4). All reflections of the new phase could be indexed in an orthorhombic cotunnite-type cell (Fig. 2). Rietveld refinement of the X-ray powder diffraction data[18] from a sample synthesized in an electrically-heated DAC (see below) at 61(2) GPa and 1100(25) K (Fig. 4) yielded atomic positions within the *Pnma* space group similar to those of $PbCl_2$ and cotunnite-type $ZrO_2$[6]. The transition from the MI to OII structure results in an increase in the coordination number of titanium atoms from 7 to 9, with oxygen atoms forming elongated tricapped trigonal prisms containing the titanium atoms.

Once synthesized at high temperature and at pressures above 60 GPa, the cotunnite-type $TiO_2$ phase could, at ambient temperature, be compressed to at least 80 GPa, decompressed to below 30 GPa, and then "recompressed" to higher pressures again (Fig. 3). Fitting the P-V data collected at ambient temperature to equation (1) gave values of $K_{300}$=431(10) GPa, $K'$=1.35(10), and $V_0$=15.82(3) cm$^3$/mol. A comparison of the molar volumes of rutile (the stable form of $TiO_2$ at ambient conditions) and the MI phase with that of the cotunnite-structured $TiO_2$ shows that the newly discovered phase is 15.9 % denser than rutile and 6.9% denser than the MI phase. This is consistent with previous observations for $ZrO_2$ and $HfO_2$[6] where the differences in the density between the phase stable at ambient conditions (MI) and the high pressure OII phase is about 13%. We also note that the bulk modulus of cotunnite-type $TiO_2$ when extrapolated to ambient conditions is just slightly lower than that of diamond.

On decompression at ambient temperature to pressures below 25 GPa, the OII phase transformed to the MI phase, and the latter transformed to $TiO_2$-II upon further decompression to between 8-12 GPa. Rapid decompression (within a second) from 60 GPa to ambient pressure in liquid nitrogen at a temperature of 77 K (using the cryogenic recovery technique similar to that described by Leinenweber et. al.[20]), however, led to the preservation of the cotunnite-structure (Fig. 2). On heating at ambient pressure to temperatures between 175 and 180 K, the quenched OII phase transformed to $TiO_2$-II.

We carried out hardness tests on the cotunnite-structured $TiO_2$ at a temperature of 155-160 K using the Vickers microhardness tester (Shimadzu Type M) with loads of 25, 50, 100, 150, and 300 g. The reliability of the hardness measurements was tested by determining the hardness of a number of polycrystalline materials ($B_4C$, $Al_2O_3$, SiC, WC, TiC, and stishovite) sintered in a DAC at 9-11 GPa and 770(25) K and subsequently quenched. The results of our measurements presented in Table 1 are in good agreement with the data in the literature. Samples of cotunnite-structured $TiO_2$ for the hardness measurements were synthesized by heating anatase or rutile to 1100(25) K at pressures of 60 to 70 GPa in an electric-heated DAC for 7 to 8 hours. It is very difficult to achieve the complete phase transition of the sample using a laser-heated DAC due to temperature gradients, particularly at the sample-diamond interface, whereas in an electrically-heated DAC, the material is heated homogeneously throughout the whole pressure chamber[21,22]. After synthesis, the samples of cotunnite-type $TiO_2$ were cryogenically recovered at 77 K. As a consequence of performing the experiment in an electrically-heated DAC, the samples are cylindrical in shape with a diameter of 250-280 $\mu$m and a thickness of 40 to 60 $\mu$m with clean, flat, surfaces suitable for hardness measurements. We conducted nine independent indentation measurements of the hardness of cotunnite-type $TiO_2$ and found the hardness to be independent of the load. All the measurements indicated a high hardness, ranging from 36.8 to 40.7 GPa, with an average value of 38 GPa.

The polycrystalline high-pressure cotunnite-structured phase of titanium dioxide is the hardest oxide discovered to date. This material is harder than stishovite and boron oxide and much harder than alumina[3,10]. Polycrystalline cubic boron nitride and sintered diamond are approximately two-times softer than their corresponding single crystal equivalents[3,7]. This suggests that the cotunnite-type $TiO_2$ is among the hardest known polycrystalline materials.

So far we have been able to preserve the cotunnite-type $TiO_2$ under ambient pressure only at low temperatures (below 170 K). By contrast, the cotunnite-type modifications of $ZrO_2$ and $HfO_2$ have been quenched to ambient conditions. As with the cotunnite-type $TiO_2$, these latter phases are found to be very incompressible[6,12]. Thus, doping of $TiO_2$ by Zr or Hf could, probably, lead to the synthesis of superhard oxides stable at ambient conditions.

*Methods*

For details of the lattice dynamic simulations see Dubrovinsky et al.[23,24]. We used the empirical Ti-O interatomic potential model by Matsui and Akaogi[25]. All calculations were performed at 300 K using a 64-point mesh in the Brillouin zone. Starting structural models for high-pressure hypothetical polymorphs were taken from Haines et al.[26]. No symmetry restrictions were used in the calculations.

We performed fully-periodic Hartree-Fock calculations as implemented in the CRYSTAL98 package[27]. We have used the TVAE* all-electron basis set, details of which are available at: http://www.dl.ac.uk/Activity/CRYSTAL. The sampling of $k$-space was performed using Pack-Monkhurst grids of shrinking factors 4, 8, and 8 yielding 11, 29, and 125 symmetry nonequivalent $k$-points for fluorite (Fm3m), pyrite ($Pa\bar{3}$), and OII structured $TiO_2$, and the values of the Coulomb and exchange truncation criteria (ITOLS parameters) were set to $10^{-7}$, $10^{-6}$, $10^{-7}$, $10^{-7}$ and $10^{-14}$ [16]. Tests revealed that increasing the number of $k$-points or improving the truncation criteria produces no significant difference in the computed structure or energy of the crystals[27]. The structural optimizations were converged to a tolerance of 0.01 Å in unit cell parameters and $10^{-5}$ Ha in the total energy using a BFGS minimization algorithm.

The full-potential linear muffin-tin-orbital (FPLMTO) calculations were based on the local-density approximation and we used the Hedin-Lundqvist parameterization for the exchange and correlation potential[24,28]. Basis functions, electron densities, and potentials were calculated without any geometrical approximation. These quantities were expanded in combinations of spherical harmonic functions (with a cut-off $l_{max} = 6$) inside non-overlapping spheres surrounding the atomic sites (muffin-tin spheres) and in a Fourier series in the interstitial region. The radial basis functions within the muffin-tin spheres are linear combinations of radial wave functions and their energy derivatives, computed at energies appropriate to their site and principal as well as orbital atomic quantum numbers, whereas outside the spheres the basis functions are combinations of Neuman or Hankel functions. In the calculations reported here, we made use of pseudo-core 3*p* and valence band 4*s*, 4*p* and 3*d* basis functions for Ti, and valence band 2*s*, 2*p*, 3*d* basis functions for O with corresponding two sets of energy parameters, one appropriate for the semi-core 3*p* states, and the other appropriate for the valence states. The resulting basis formed a single, fully hybridizing basis set. For sampling the irreducible wedge of the Brillouin-zone we used the special k-point method[19].

The details of experiments in electrically- and laser-heated DAC are described in our earlier works[21,22,24,29,30]. We obtained powder X-ray diffraction data with a Siemens X-ray system consisting of a Smart CCD Area Detector and a direct-drive rotating anode X-ray generator (18 kW). Mo$K_\alpha$ radiation (tube voltage 50 kV, tube current 24 mA) focused with a capillary X-ray optical system to ⌀35 μm FWHM was used. The Rietveld refinements of powder X-ray diffraction data were carried out using the GSAS program[18]. In the studies of EOS (Fig. 3) of different $TiO_2$ polymorphs Pt or Au powders mixed in the mass proportion approximately 20:1 with the sample were used as internal pressure standard for X-ray diffraction and ruby as pressure marker. Experiments with MI phase have been conducted in Ar pressure medium. In experiments on EOS with OII phase after each pressure increases or decreases stresses in the sample was relaxed by external electrical heating of the cell at 800-850 K during 1 hour. Only data on OII synthesized in electrically-heated DACs were used due to uncertainty in chemical composition of laser-heated samples. Diffraction pattern shown on Fig. 2, 4 were collected for pure $TiO_2$ and pressures were measured from predetermined EOS of corresponding phases of $TiO_2$.

**Figure Captions**

Figure 1. The stabilities of various known and hypothetical $TiO_2$ polymorphs relative to rutile as function of pressure obtained by lattice dynamics (LD) at T=300 K (a) and by *ab initio* periodic Hartree-Fock calculations at T=0 K (b). The designation of the phases follows Ref. 6: $TiO_2$-II has the $\alpha$-$PbO_2$ structure, MI has baddeleyite structure, OI is orthorhombic with *Pbca* space group, and OII is a cotunnite-type structure. (c) The internal energy (for 2 $TiO_2$ formula units) with respect to volume for a selection of high-pressure $TiO_2$ polymorphs as computed using *ab initio* theory. In LD simulations the MI phase was traced only to 30 GPa and OII only to 40 GPa because at lower pressures the MI and OII phases relaxed to the $TiO_2$-II structure.

Figure 2. Examples of X-ray diffraction patterns obtained in experiments with titanium dioxide. (a) Anatase at 6.5(3) GPa and 290 K ($a$=3.7585(6) Å, $c$=9.3500(11) Å). (b) Baddeleyite-type (MI) phase at 27(1) GPa and 290 K ($a$= 4.573(1) Å, $b$= 4.820(2) Å, $c$= 4.769(2) Å, $\beta$=98.6(1)). (c) Cotunnite-type (OII) phase obtained by laser-heating anatase compressed to 64(2) GPa and heated at 1800(100) K for 40 min. The diffraction data collected from a sample temperature-quenched to 290 K ($a$=5.154(2) Å, $b$=2.984(2) Å, $c$=5.951(3) Å). (d) OII phase synthesized in the electrically-heated DAC by heating at 1100(25) K for 8 hours at a pressure of 70(2) GPa. After the synthesis, the sample was cryogenically recovered in liquid nitrogen and the diffraction pattern collected at ambient pressure and 160(2) K ($a$=5.246(3) Å, $b$=3.270(3) Å, $c$=6.126(4) Å). Backgrounds were subtracted from all the spectra.

Figure 3. Pressure dependence of volume of anatase, baddeleyite-type, and cotunnite-type phases of $TiO_2$. Birch-Murnaghan equations of state (equation (1)) are plotted as solid lines with parameters $K_{300}$=178(1) GPa, K'=4 (fixed), and $V_0$=20.59(1) cm$^3$/mol for anatase; $K_{300}$=304(6) GPa, K'=3.9(2), and $V_0$=16.90(3) cm$^3$/mol for the MI phase; and $K_{300}$=431(10) GPa, K'=1.35(10), and $V_0$=15.82(3) cm$^3$/mol for the cotunnite-type (OII) phase.

Figure 4. An example of profile-fitted X-ray diffraction data obtained from a cotunnite-structured $TiO_2$ sample at 290 K. The sample was synthesized in the electrically-heated DAC at 61(2) GPa and 1100(25) K and subsequently temperature quenched to 290 K. The GSAS program package[18] was used in the Rietveld refinement ($wR_p$=1.9%, $R_p$=1.6%, $\chi^2$=0.49).

Table 1. Bulk moduli and Vickers hardness for some polycrystalline hard materials.

| Material | Bulk modulus, GPa | Hardness, GPa | Source |
|---|---|---|---|
| $B_4C$ | 200 | 30 | 3 |
| $B_4C$ | 200 | 30(2) | This study |
| $B_6O$ | 228 | 30 | 10 |
| TiC | 241 | 29 | 8 |
| TiC | 241 | 30(3) | This study |
| SiC | 248 | 29 | 3 |
| SiC | 248 | 29(3) | This study |
| $Al_2O_3$ | 252 | 19 | 7 |
| $Al_2O_3$ | 252 | 20(2) | This study |
| $SiO_2$, stishovite | 291 | 33 | 7 |
| $SiO_2$, stishovite | 291 | 32(2) | This study |
| WC | 421 | 30 | 20 |
| WC | 421 | 30(3) | This study |
| Cubic BN | 369 | 32 | 3 |
| Cottunite-type $TiO_2$* | 431 | 38(3) | This study |
| Sintered diamond | 444 | 50 | 3 |

*Measurements at 157(2) K

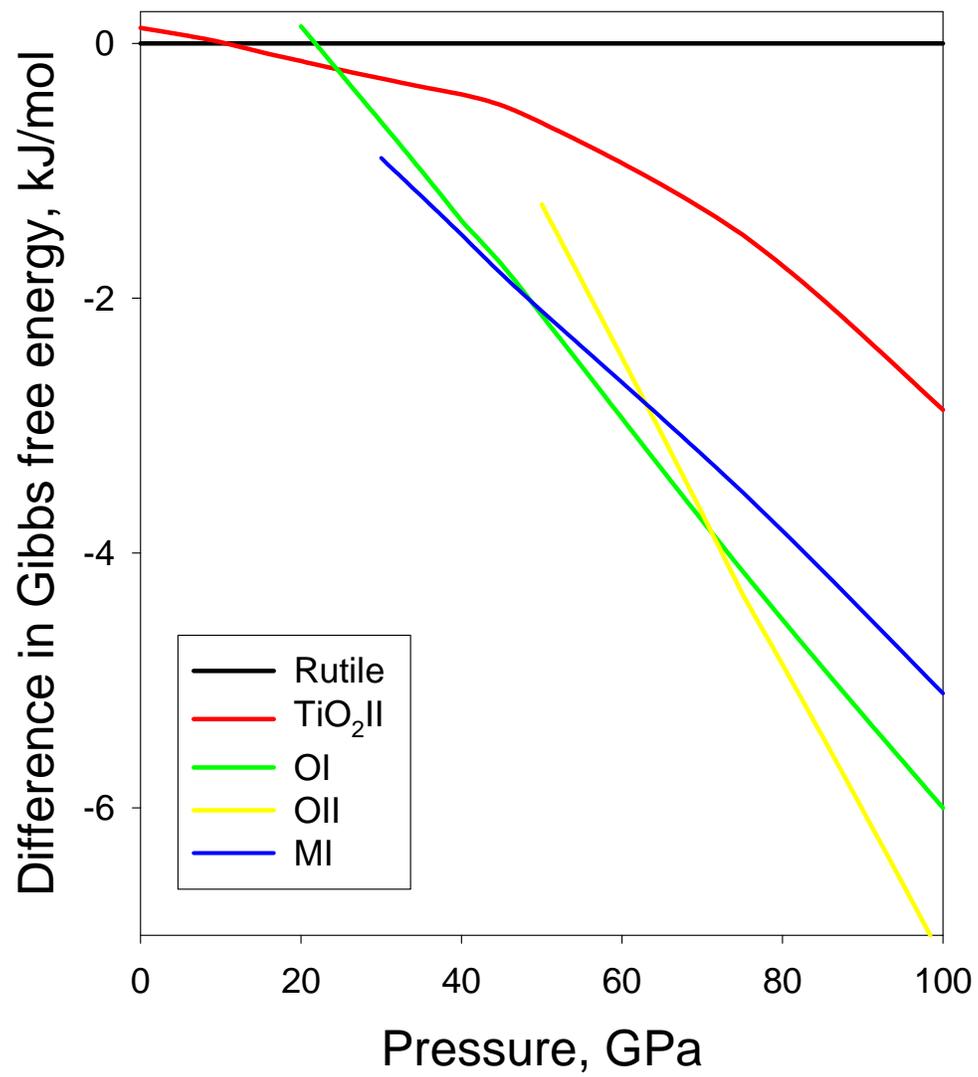

Fig. 1a.

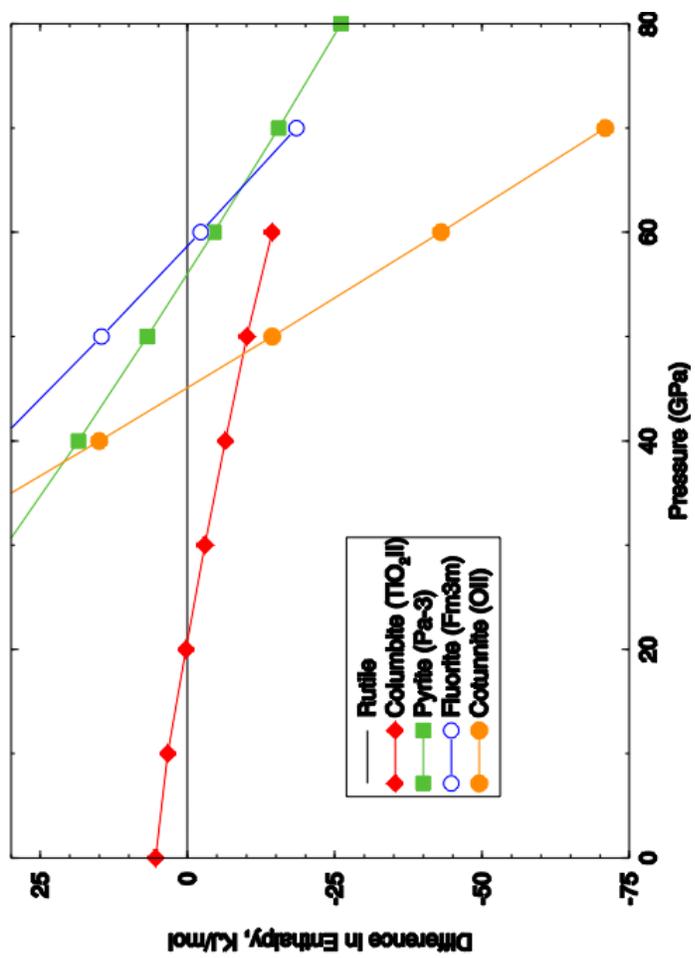

Fig. 1b.

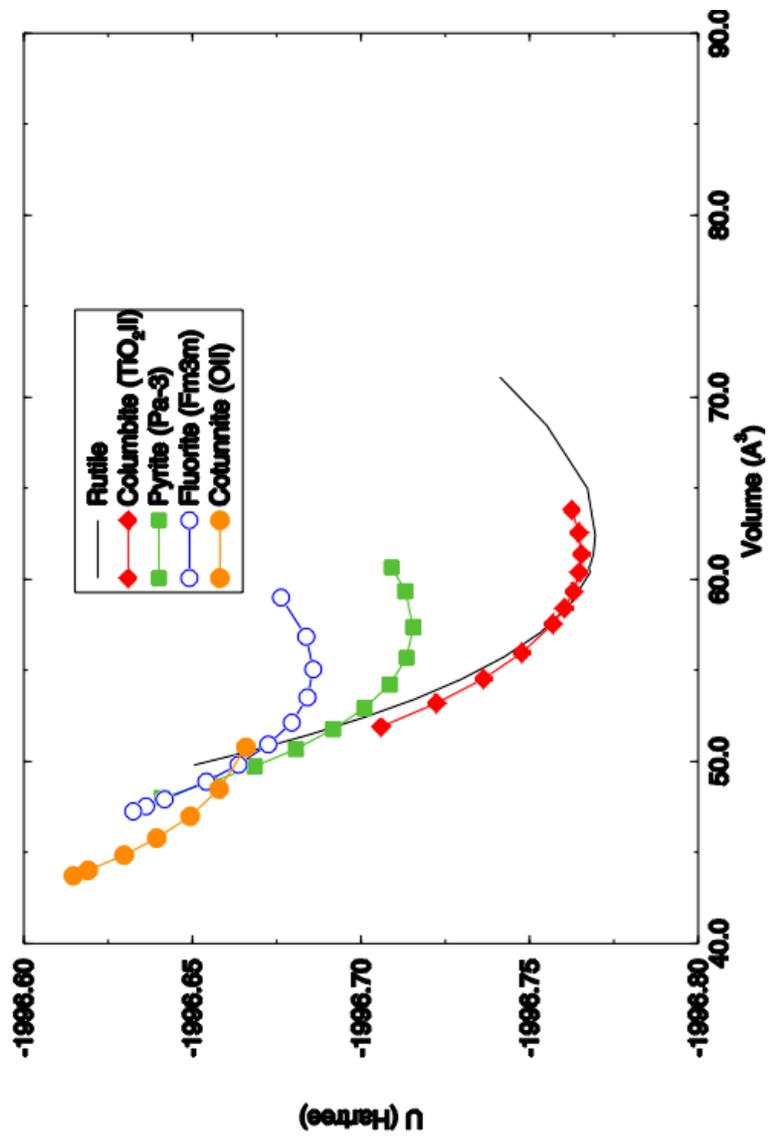

Fig. 1c.

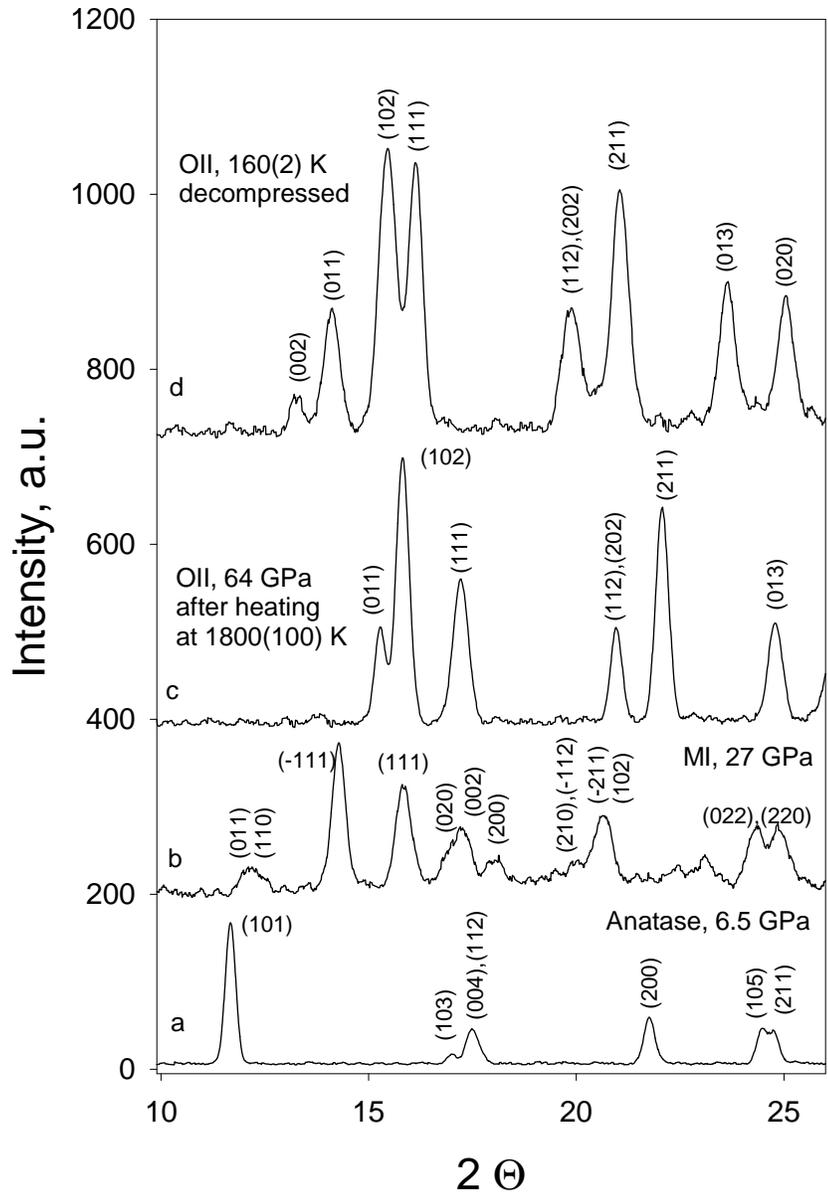

Fig. 2.

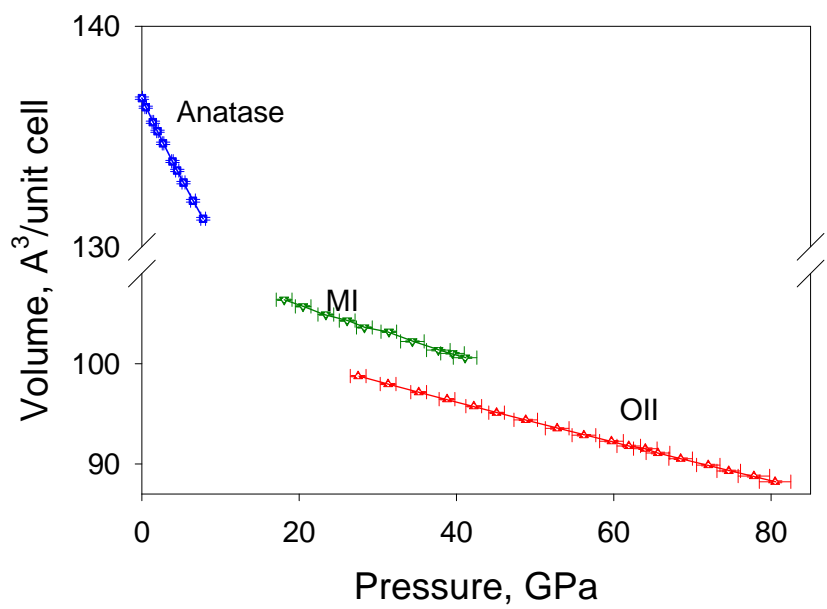

Fig. 3.

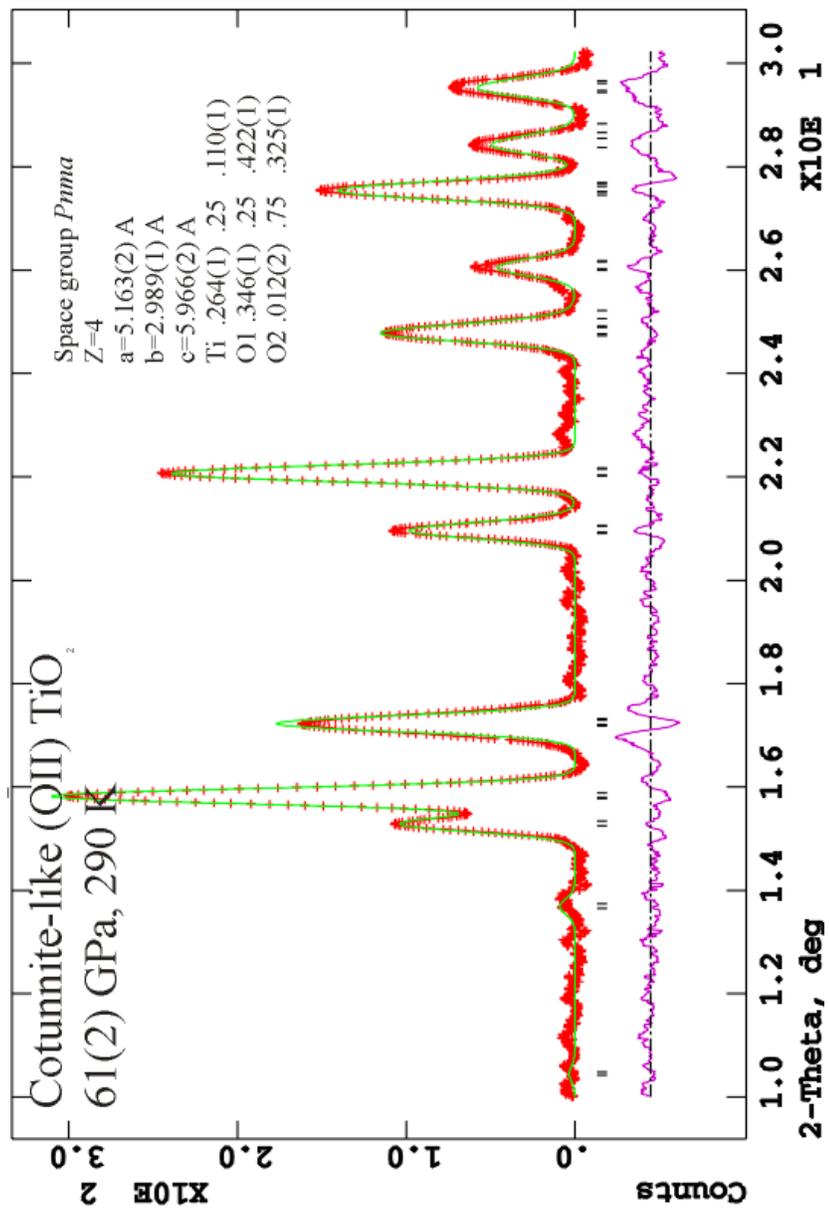

Fig. 4.